\documentclass[10pt]{article}
\usepackage[cp1251]{inputenc}
\usepackage[dvips]{graphicx}
\def\lsim{\
  \lower-1.2pt\vbox{\hbox{\rlap{$<$}\lower5pt\vbox{\hbox{$\sim$}}}}\ }
\def\gsim{\
  \lower-1.2pt\vbox{\hbox{\rlap{$>$}\lower5pt\vbox{\hbox{$\sim$}}}}\ }

\usepackage{caption2}

\begin{document}
\title{Calculation of the one-particle and two-particle
condensates in He-II at $T=0$}
\author{ \textbf{Maksim Tomchenko}
\bigskip \\ {\small Bogolyubov Institute for Theoretical Physics,}
  \\ {\small Metrologichna St. 14-b, 03143 Kiev, Ukraine}
   \\{\small E-mail: mtomchenko@bitp.kiev.ua}}
 \date{\empty}
 \maketitle
 \large
 \sloppy

 \textsl{We analyze  the microstructure of He-II in the framework
 of the method of collective  variables (CV), which was proposed
 by Bogolyubov and Zubarev and was developed later by
 Yukhnovskii and Vakarchuk. The logarithm of the ground-state
 wave function of He-II, $\ln{\Psi_{0}}$, is calculated in the
  approximation of ``two sums'', i.e., as a Jastrow function and first
 (three-particle) correction. In the CV method equations for $\Psi_{0}$
 are deduced  from the $N$-particle Schr\"{o}dinger
 equation.  We also take into account the connection between
 the structure factor and $\Psi_0$, which allows one to obtain
$\Psi_{0}$ from the structure factor of He-II, not from a model
 potential of interaction between He-II atoms.
 It should  be emphasized that the model does not have any free
 parameters or functions. The amount of one-particle ($N_1$) and two-particle
 ($N_2$) condensates is calculated for the ground state
  of He-II: we find $N_1\approx 0.27N$ and $N_2\approx 0.53N$
  in the Jastrow approximation for $\Psi_{0}$, and, taking into account
the three-particle correction to $\ln{\Psi_{0}}$,
 we obtain $N_1\approx 0.06N$ (which agrees with the experiment)
 and   $N_2\approx 0.16N$. In the   approximation
 of ``two sums'', we  also find that
 the higher $s$-particle condensates ($s\geq 3$) are absent in He-II at $T=0$.}

\bigskip \\

             \section{Introduction}
 The sum of one-particle condensate (1PC), two-particle  condensate (2PC)
 and  higher $s$-particle
  condensates is usually referred to as the composite condensate.
  Knowing the structure of the composite condensate
  in He-II is, without doubt, of great importance
  \cite{bog}--\cite{pash}.    Except for a purely
 cognitive interest, it also has a ``practical'' side, namely, in the
 field-theoretic approaches to the modelling of He-II
 microstructure \cite{bog,bruck,2ch,pash}, the
 quasiparticle spectrum of He-II is explicitly expressed through
 the amount of the 1PC and 2PC, and the dependence on higher
 condensates is also not excluded. Of interest is the question
 whether all of He-II atoms belong to the composite condensate at
   $T=0$, as was suggested in \cite{shev,nep,pash}. The superfluidity
 of He-II by itself is probably \cite{yang,odlro} caused by an
 off-diagonal long-range order (ODLRO).

Our work is devoted to a calculation of the  amount of 1PC and 2PC in He-II at
$T=0$. To describe the microstructure of He-II, we use the method  of
collective  variables (CV), which was first proposed
 by Bogolyubov and Zubarev \cite{bz} and was later developed in the works
 by  Yukhnovskii and Vakarchuk \cite{yuvdan}--\cite{vak2}. First, we obtain the ground-state
 wave function of He-II, $\Psi_{0}$, and then we calculate the amount
 of the condensates using the formula of \cite{vak1,fnt} for the $s$-particle density
 matrices $F_{s}$.
 The model does not contain any free parameters or functions:
 $\Psi_{0}$ is obtained as an eigenfunction of  the $N$-particle Schr\"{o}dinger
 equation; we also take into account the connection between $\Psi_{0}$
  and the structure factor $S(k)$ of He-II \cite{vak1,fnt}.

  As far as we know, the higher $s$-particle condensates ($s\geq 3$) were not
  calculated previously. The amount of 1PC in He-II at $T=0$ was found in many works
  \cite{ponz,yuvdan,vak2,man}--\cite{mor,pash} (though only in \cite{yuvdan,vak2,qdmc,mor}
  it was done without free parameters), and the theory agrees with the
  experiment on the whole. The amount of 2PC in He-II is still unknown.
 The 1PC was measured in many works, as a number of
atoms with momentum equal to zero, but how to measure the 2PC is
not yet clear (this issue is beyond of the scope of the article).
 There are only several theoretical estimates
  of the 2PC \cite{ris,nep}, all of which use free parameters.
  The results of \cite{ris} are discussed below in Sec.~4.
   In  \cite{nep}, a final result was not presented, but some equations were
   obtained, from which it follows that the proportion
   between 1PC and 2PC can vary depending on the form of the potential of
   the interaction between He$^4$ atoms. We calculate the amounts of 1PC and 2PC
   without free parameters, and in more exact  approximation as compared with
   \cite{yuvdan,vak2,ris}.

   In \cite{yuvdan} and \cite{vak2}, the amount of 1PC was found in the
    approximation of ``one sum'' (1S) with the result $N_{1} \approx 0.08N$
    and $N_{1} \approx 0.04N$, respectively. Below, we obtain 1PC ($N_{1}$)
    and 2PC ($N_{2}$) for He-II at $T=0$ in more
   exact  approximation  of ``two sums'' (2S), and we find
    $N_{1} \approx 0.06N$ and $N_{2} \approx 0.16N$.
   Our formula for 2PC refines the Ristig's formula
 \cite{ris} previously obtained by another method in the 1S-approximation for
 $\ln{\Psi_{0}}$.

   In \cite{fnt} we have found also that (i)~in a weakly interacting
  Bose gas, all atoms belong to 1PC or 2PC at $T=0$, and  (ii)~the
   higher $s$-particle condensates ($s\geq 3$) are absent in He-II at $T=0$
   (which was shown in the 2S-approximation).

          \section{The ground-state  wave function of He-II}
There exist several methods for calculation of the ground-state
wave function of He-II  (see also the review  \cite{obz}): the
variational method \cite{woo}, the ``Green's function Monte
Carlo'' method \cite{gfmc} and it's development, ``shadow wave
function'' (SWF) method \cite{swf},  the Path Integral Monte Carlo
simulations \cite{pimc}, the  diffusion Monte Carlo (MC)
simulations \cite{qdmc,mor}, the ``hypernetted chain'' (HNC)
method \cite{hnc,hnc2}, (all of these are indirect methods for
solving the $N$-particle Schr\"{o}dinger equation), Feenberg's
approach \cite{feenb} (solving the Schr\"{o}dinger equation in the
${\bf r}$-space), and the CV-method \cite{yuvdan}--\cite{fnt},
\cite{ujp,megot} (solving the Schr\"{o}dinger equation in the
${\bf k}$-space).

   The quantum-mechanical models are developed actively
and a noticeable  progress is already achieved
\cite{fnt,ujp,mor,obz,swf,hnc2,gl}. Modern variational methods
\cite{swf} (SWF), \cite{hnc,hnc2} (HNC) reach an accuracy of the
order better then $0.1\,K$, but such models, unfortunately, use
several free parameters.

 In the CV-method,  the equations are deduced from the
first principles, namely, from the exact $N$-particle
Schr\"{o}dinger equation, and a solution of these equations can be
found numerically without using any fitting parameters or
functions (which is important) by taking into account the
connection between $\Psi_{0}$ and $S(k)$.

All existing models of He-II of which we are aware (except the
mentioned CV-approach) use several fitting parameters, at least,
in the effective interaction potential (even the MC simulations);
in this case, of course, it is not so difficult to obtain two
``points'': the ground-state energy and 1PC. In our opinion, the
approach without fitting parameters is preferable because, in this
case, nothing is introduced in the model ``by hand''.  In our
paper, we use the CV-method since this method
  does not have any fitting parameters.

According to \cite{yuv1}, the ground-state  wave function of
 He-II has the form
 \begin{equation}
  \Psi_0 = \frac{e^{{\small S_{0}}}}{\sqrt{Q}}, \quad
   S_{0} =\sum\limits_{j\geq 2}\frac{N^{1-j/2}}{j!}
   \sum\limits_{{\bf k}_{1},\ldots,{\bf k}_{j}\not= 0}
  \delta({\bf k}_{1}+\ldots+{\bf k}_{j})a_{j}({\bf k}_{1},\ldots,{\bf k}_{j})
   \rho_{{\bf k_{1}}}\ldots\rho_{{\bf k_{j}}},
    \label{osn}     \end{equation}
 where $\delta$ is the Kronecker delta, $N$ is the total number of atoms in helium,
 $Q$ is a normalization constant, and $\rho_{{\bf k}}$ are  the collective
 variables
     \begin{equation}
   \rho_{{\bf k}} = \frac{1}{\sqrt{N}}
 \sum\limits_{j=1}^{N}e^{-i{\bf k}{\bf r}_j} , \quad {\bf k}\not= 0.
 \label{rok}    \end{equation}
The term with  $a_n$ in (\ref{osn}) corresponds to  $n$-particle correlations
in ${\bf r}$-space. The zero-order approximation for $\Psi_{0}$, which is also
the 1S-approximation, is
  \begin{equation}
  \ln{\Psi_0} = -\frac{1}{2}\ln{Q}+
  \sum\limits_{{\bf k}\not= 0}\frac{a_{2}(k)}{2}
     \rho_{{\bf k}}\rho_{-{\bf k}}, \quad a_{n\geq 3} = 0.
    \label{osnj}     \end{equation}
The wave function $\Psi_{0}$ (\ref{osnj}) can be present in the well-known
Jastrow form
\begin{equation}
  \Psi_0 = \frac{1}{\sqrt{Q}}\prod\limits_{i,j}
  e^{S_1({\bf r_{i}}-{\bf r_{j}})},
    \label{jast}     \end{equation}
 where
    \begin{equation}
 S_1(r) = \frac{1}{N}\sum\limits_{{\bf k}}\frac{a_{2}(k)}{2}
 e^{i{\bf k}{\bf r}}=\frac{V}{N}\frac{1}{(2\pi)^{3}}\int d{\bf k}
 \frac{a_{2}(k)}{2}e^{i{\bf k}{\bf r}},
   \label{S11}    \end{equation}
and $V$ is the volume of the system.

The capabilities of computers force us to restrict ourselves to the
2S-approximation for $\Psi_{0}$; in this case, the sums with $a_{2}$ and
$a_{3}$ are taken into account in (\ref{osn}), but $a_{n\geq 4} = 0$ (for the
``$n$-sum'' approximation \cite{vak1,vak2} all sums up to the $n$-th sum over
${\bf k}$ are included in all expansions). Substitution of $\Psi_{0}$
(\ref{osn}) into the $N$-particle Schr\"{o}dinger equation allows one to obtain
\cite{yuv1} a chain of equations for $a_{n}$. In the 2S-approximation, we have
 \begin{equation}
  a_{3}({\bf k}_{1},{\bf k}_{2}) = -\frac{2a_{2}(k_{1})a_{2}(k_{2})
  {\bf k}_{1}{\bf k}_{2} + 2a_{2}(k_{1})a_{2}(k_{3})
  {\bf k}_{1}{\bf k}_{3} + 2a_{2}(k_{2})a_{2}(k_{3})
  {\bf k}_{2}{\bf k}_{3}}{k_{1}^{2}[1-2a_{2}(k_{1})] +
  k_{2}^{2}[1-2a_{2}(k_{2})] + k_{3}^{2}[1-2a_{2}(k_{3})]}.
       \label{a3} \end{equation}
   Throughout in the paper, we assume ${\bf k}_{3}=-{\bf
k}_{1}-{\bf k}_{2} \not = 0$. The $s$-particle density matrices $F_{s}$  for
the ground-state of Bose liquid  were found in \cite{vak1} for the
approximation of 1S, and in \cite{vak1,fnt} for the approximation of 2S .
Moreover, in \cite{vak1,fnt}, an equation connecting $\Psi_{0}$ with the
structure factor $S(k)$ was obtained:
 \begin{equation}
  2a_{2}(k) = 1-\frac{1}{S(k)}-\frac{\Sigma_{1}(k)}{S(k)[1-2a_{2}(k)]}-
   \frac{\Sigma_{2}(k)}{S(k)},
   \label{a2}  \end{equation}
  where
  \begin{equation}
  \Sigma_{1}(k) = \frac{2}{N}\sum\limits_{{\bf q}\not = 0}
  \frac{a_{2}(q)a_{2}({\bf k}+{\bf q}) + a_{3}({\bf k},{\bf q})
  \left [1+a_{3}({\bf k},{\bf q}) \right ]}{[1-2a_{2}(q)]
  [1-2a_{2}({\bf k}+{\bf q})]},
   \label{sig1}  \end{equation}
\begin{equation}
  \Sigma_{2}(k) = \frac{2}{N}\sum\limits_{{\bf q}\not = 0}
  \frac{a_{3}^{2}({\bf k},{\bf q})}
{[1-2a_{2}(q)][1-2a_{2}({\bf k}+{\bf q})]},
   \label{sig2}  \end{equation}
and also the formula for the amount of 1PC ($N_1$) was found:
\begin{equation}
 \ln{\left (N_{1}/N\right )}=I_{1A}+I_{2A}+I_{2B}+I_{2C},
 \label{N0}  \end{equation}
\begin{equation}
  I_{1A} = -\frac{1}{N}\sum\limits_{{\bf k}\not = 0}
  \frac{a_{2}^{2}(k)}{[1-2a_{2}(k)]},
   \label{I1A}  \end{equation}
\begin{equation}
  I_{2A} = -\frac{1}{8N^{2}}\sum\limits_{{\bf k}_{1},{\bf k}_{2}\not = 0}
 \left (\prod\limits_{j=1}^{3}\frac{2a_{2}(k_{j})}{1-2a_{2}(k_{j})} \right )
  \frac{1}{1-2a_{2}(k_{1})},
   \label{I2A}  \end{equation}
\begin{eqnarray}
  I_{2B} &=& -\frac{1}{2N^{2}}\sum\limits_{{\bf k}_{1},{\bf k}_{2}\not = 0}
 \left (\prod\limits_{j=1}^{3}\frac{1}{1-2a_{2}(k_{j})} \right )
 a_{3}({\bf k}_{1},{\bf k}_{2})*
 \nonumber \\ &*&\left \{\frac{2a_{2}(k_{1})+
 a_{3}({\bf k}_{1},{\bf k}_{2})}{1-2a_{2}(k_{1})}
  + 2a_{2}(k_{1})[1-a_{2}(k_{2})] \right\},
   \label{I2B}  \end{eqnarray}
 \begin{equation}
  I_{2C} = \frac{1}{2N^{2}}\sum\limits_{{\bf k}_{1},{\bf k}_{2}\not = 0}
   a_{3}^{2}({\bf k}_{1},{\bf k}_{2})
 \prod\limits_{j=1}^{3}\frac{1}{1-2a_{2}(k_{j})}.
   \label{I2C}  \end{equation}

 Note that equations similar to (\ref{a3}), (\ref{a2}) were
 obtained by other methods by Campbell and Krotschek \cite{hnc}.
 Our system of Eqs. (\ref{a3})--(\ref{sig2}) is a little more
 exact, because in (\ref{a3}) we have $a_3$ with $a_2$ from
 (\ref{a2}) (which includes correction $a_3 \neq 0$), but in
  \cite{hnc} $a_3$ was taken with $a_2$ in zeroth approximation
  ($a_3 = 0$ in (\ref{a2})).

The following asymptotics as $k\rightarrow 0$ is true \cite{fnt}:
  \begin{equation}
 2a_{2}(k\rightarrow 0) = -\frac{1+\Sigma_{2}(0)}{S(k)},
   \label{a20} \end{equation}
   \begin{equation}
 E(k\rightarrow 0) = ck = \frac{\hbar^{2}k^{2}}{2mS(k)}
  \left (1+\Sigma_{2}(0) \right ),
   \label{E0} \end{equation}
     \begin{equation}
  S(k\rightarrow 0) = \frac{\hbar k}{2mc}(1+\Sigma_{2}(0)),
    \label{Sk} \end{equation}
where $E(k)$ is the quasiparticle spectrum of Bose liquid (see the equations
for $E(k)$ in \cite{yuv2,jetp,megot}), and
 \begin{equation}
  \Sigma_{2}(0) = \frac{8}{N}\sum\limits_{{\bf q}\not = 0}
  \left [\frac{a_{2}(q)}{1-2a_{2}(q)} \right ]^4 > 0.
  \label{sig20} \end{equation}
In the 1S-approximation (\ref{osnj}), for which
 \begin{equation}
  2a_{2}(k) =1-\frac{1}{S(k)}, \quad a_{n\geq 3}=0,
     \label{a2o}     \end{equation}
for He-II  we obtain $\Sigma_{2}(0)=0.33$ (from (\ref{sig20},\ref{a2o})), and,
in the 2S-approximation, a numerical solution of (\ref{a3})--(\ref{sig2}) (see
Sec.~3) gives $\Sigma_{2}(0)=0.66$.

   \section{One-particle condensate in  He-II at $T=0$}

  In the 1S-approximation, the amount of 1PC can be simply found
  from  (\ref{N0}), (\ref{I1A}), and (\ref{a2o}). For the structure factor
  $S(k)$,
  we use the smoothed experimental data  from \cite{ss}, which is, perhaps,
  the most exact. We extrapolate this data to
  $T=0$ according to \cite{fs}
\begin{equation}
  S(k,T\!=\!0) = S(k,T)\tanh{\frac{E(k)}{2k_{B}T}}.
   \label{ST0}  \end{equation}
At  $T=0$, we should have the asymptotic $S(k\!=\!0)=0$ \cite{bal};
 in (\ref{ST0}), we take into account that $E(k\rightarrow 0)=ck$.
 From (\ref{N0}), (\ref{I1A}), (\ref{a2o}), and (\ref{ST0}), we numerically
 find
$n_{1}\equiv \frac{N_{1}}{N}\cdot100\% = 27.2\%$.

 To calculate $N_1$ in the 2S-approximation, we need to know, according to
 (\ref{N0})--(\ref{I2C}), the functions $a_{2}(k)$  and $a_{3}({\bf k}_{1},{\bf k}_{2})$.
 The function $a_{3}$ is defined in (\ref{a3}), and to obtain $a_{2}(k)$
  we should solve numerically the integral equation (\ref{a2})
  taking into account (\ref{a3}), (\ref{sig1}), and (\ref{sig2}). We everywhere
  replace sums by integrals according to the rule
  \cite{yuv1,vak1}
  \begin{equation}
\sum\limits_{{\bf k}}\rightarrow\frac{V}{(2\pi)^{3}}\int d{\bf k}.
     \label{sum}     \end{equation}
Equation  (\ref{a2}) cannot be solved by iteration; we have
succeeded in solving it numerically by the Newton's method, in
which way we found $a_{2}(k)$. From (\ref{a3}),
(\ref{N0})--(\ref{I2C}), and (\ref{sum}), we obtain $n_{1} =
6.1\%$, which agrees with the experiment: $n^{exp}_{1} \approx
6-12 \%$ for $T=0$ \cite{blag}--\cite{new}. We estimate the
numerical error for $n_{1}$ to be $\frac{\delta n_{1}}{n_1} \simeq
0.1$. The properties of $\Psi_{0}$ and the possibility to obtain
the ground-state energy $E_0$ are discussed in more detail in
\cite{ujp}.

    \section{Two-particle condensate}
 It can be shown in 2S-approximation that the $s$-particle
 density matrices $F_{s}$ of helium-II for $T=0$ in three dimensions display
 ODLRO:
  \begin{equation}
   \lim\limits_{\vert {\bf r}_i-{\bf r}_j^{\prime}\vert\rightarrow
  \infty} F_{s}({\bf r}_1,\ldots,{\bf r}_s\vert{\bf r}_1^{\prime},\ldots,{\bf
  r}_s^{\prime}){\large \vert}_{B} =F_{s}(\infty)=\left [F_{1}(\infty)\right ]^s = \mbox{const}>0,
     \label{Fsinf}     \end{equation}
     \[ \mbox{B:} \quad
   \vert {\bf r}_i-{\bf r}_j\vert, \
 \vert {\bf r}_i^{\prime}-{\bf r}_j^{\prime}\vert
 \quad\mbox{are fixed for any} \quad  i,j=1,\ldots,s,\quad i\neq j. \]
   From the general principles, we write  the probability $W_{{\bf k}_{1},{\bf k}_{2}}$
   of finding the momenta ${\bf k}_1$  and  ${\bf k}_2$ in two arbitrary atoms, for $T=0$:
 \begin{eqnarray}
&W_{{\bf k}_{1},{\bf k}_{2}}& = \frac{1}{V^{2}}\int d{\bf
r}_{3}\ldots d{\bf r}_{N}
 \left \vert \int\Psi_{0}({\bf r}_{1},\ldots,{\bf r}_{N})e^{i{\bf k}_{1}{\bf r}_{1}+
 i{\bf k}_{2}{\bf r}_{2}}d{\bf r}_{1}d{\bf r}_{2}\right\vert^{2}
 = \nonumber\\ &=&
\frac{1}{V^{4}}\int F_{2}({\bf r}_{1},{\bf r}_{2}\vert {\bf
r}_{1}^{\prime},{\bf r}_2^{\prime}) e^{i{\bf k}_{1}({\bf r}_{1}-
{\bf r}_{1}^{\prime}) + i{\bf k}_{2}({\bf r}_{2}-{\bf
r}_{2}^{\prime})} d{\bf r}_{1}d{\bf r}_{2}d{\bf
r}_{1}^{\prime}d{\bf r}_{2}^{\prime},
  \label{w12}     \end{eqnarray}
  where $F_2$ is the two-particle density matrix.

  To obtain the amount of 2PC, we should know $W_{{\bf k},-{\bf k}}$ for $T=0$.
 Let us consider first the 1S-approximation for $\Psi_{0}$ and $F_2$.
 In this approximation,  $\Psi_{0}$ is defined by (\ref{osnj}), and  $F_{s}=F_{s}^{(1)}$
was found in \cite{vak1}:
   \begin{eqnarray}
  F_{s}^{(1)}({\bf r}_1,\ldots,{\bf r}_s\vert{\bf r}_1^{\prime},\ldots,{\bf
  r}_s^{\prime}) &=& exp\left \{\sum\limits_{{\bf k}\neq 0} \left
  [\frac{f_{1}(k)}{2}\left (|\xi_{{\bf k}}|^2 + |\xi^{\prime}_{{\bf k}}|^2
  \right )\right.\right. - \nonumber\\ &-& \left.\left.
    \frac{s}{N}\frac{a_{2}(k)}{1-2a_{2}(k)} +
   f_{2}(k)\xi_{{\bf k}}\xi^{\prime}_{-{\bf k}} \right ]
   \right\} ,
   \label{Fs1}     \end{eqnarray}
   where
 \begin{equation}
   f_{1}(k)=a_{2}(k)+f_{2}(k), \quad
   f_{2}(k)=\frac{a_{2}^{2}(k)}{1-2a_{2}(k)},
     \label{fi}     \end{equation}
 \begin{equation}
   \xi_{{\bf k}}=\frac{1}{\sqrt{N}}\sum\limits_{j=1}^{s}
   e^{-i{\bf k}{\bf r}_{j}},  \quad
   \xi^{\prime}_{{\bf k}}=\frac{1}{\sqrt{N}}\sum\limits_{j=1}^{s}
   e^{-i{\bf k}{\bf r}^{\prime}_{j}}.
     \label{ksi}     \end{equation}
  It is convenient to represent  $F_{2}^{(1)}$ in the form
     \begin{eqnarray}
  && \ln{F_{2}^{(1)}}=\ln{F_{2}(\infty)} +
  \varphi_{1}({\bf r}_{1}-{\bf r}_{2}) +
  \varphi_{1}({\bf r}^{\prime}_{1}-{\bf r}^{\prime}_{2}) + \nonumber\\
 &+& \varphi_{2}({\bf r}^{\prime}_{1}-{\bf r}_{1}) +
   \varphi_{2}({\bf r}^{\prime}_{2}-{\bf r}_{2}) +
   \varphi_{2}({\bf r}^{\prime}_{2}-{\bf r}_{1}) +
   \varphi_{2}({\bf r}^{\prime}_{1}-{\bf r}_{2}),
  \label{F2new}     \end{eqnarray}
   \begin{equation}
  \varphi_{i}({\bf r})=\frac{1}{N}\sum\limits_{{\bf q}\neq 0}
  f_{i}({\bf q})e^{-i{\bf q}{\bf r}}.
     \label{phi}     \end{equation}
 The functions  $\varphi_{i}$ have the property $\varphi_{i}(r\rightarrow\infty)\sim
 \frac{1}{r^2}\rightarrow 0$. From (\ref{w12}), we obtain
   \begin{eqnarray}
  & W_{{\bf k},-{\bf k}}& = \frac{F_{2}(\infty)}{V^{4}}\int
   d{\bf r}_{1}d{\bf r}_{2}d{\bf r}_{1}^{\prime}d{\bf
r}_{2}^{\prime} e^{i{\bf k}({\bf r}_{1}- {\bf r}_1^{\prime}) - i{\bf k}({\bf
r}_{2}-{\bf r}_2^{\prime})} \exp \left [
    \varphi_{1}({\bf r}_{1}-{\bf r}_{2}) \right. +  \label{w2}
    \\ &+&
 \left. \varphi_{1}({\bf r}^{\prime}_{1}-{\bf r}^{\prime}_{2}) +
  \varphi_{2}({\bf r}^{\prime}_{1}-{\bf r}_{1}) +
   \varphi_{2}({\bf r}^{\prime}_{2}-{\bf r}_{2}) +
   \varphi_{2}({\bf r}^{\prime}_{2}-{\bf r}_{1}) +
   \varphi_{2}({\bf r}^{\prime}_{1}-{\bf r}_{2})\right ].
   \nonumber
      \end{eqnarray}
    To calculate $W_{{\bf k},-{\bf k}}$  (\ref{w2}), we expand
    the exponent in (\ref{w2}), with the sum of all $\varphi_{i}$, in a power series.
    It can be shown, that only integrals of terms of the form
        \begin{equation}
  \varphi^{l_{1}}_{1}({\bf r}_{1}-{\bf r}_{2})
  \varphi^{l_{2}}_{1}({\bf r}^{\prime}_{1}-{\bf r}^{\prime}_{2})
  \quad  \mbox{and} \quad
  \varphi^{l_{3}}_{2}({\bf r}^{\prime}_{1}-{\bf r}_{1})
   \varphi^{l_{4}}_{2}({\bf r}^{\prime}_{2}-{\bf r}_{2})
    \label{slag}     \end{equation}
  ($l_j= 1,2,3,\ldots$ are all possible natural numbers)
  are significant for the value of $W_{{\bf k},-{\bf k}}$.
  The integrals of other sets of $\varphi_{i}$ are $N$ times
  smaller or equal to zero. Taking all this into account, we obtain
  \begin{equation}
  W_{{\bf k},-{\bf k}} = \frac{F_{2}(\infty)}{N^{2}}
  \left [\Phi_{1}^{2}(k) + \Phi_{2}^{2}(k) \right ], \quad k\neq 0,
     \label{w22}     \end{equation}
     \begin{equation}
  \Phi_{i}(k) = \frac{N}{V}\int
  e^{\varphi_{i}(R)}e^{i{\bf k}{\bf R}}d{\bf R},
  \quad F_{2}(\infty)=F_{1}^2(\infty).
       \label{Phi}     \end{equation}
  Following the notation of \cite{ris}, we rewrite $W_{{\bf k},-{\bf k}}$
  (\ref{w22}) in the form
   \begin{equation}
  N^{2}W_{{\bf k},-{\bf k}} = N_{{\bf k}}N_{-{\bf k}} +
   \chi_{{\bf k}}\chi_{-{\bf k}},  \quad k\neq 0,
     \label{wris}     \end{equation}
     where
  \begin{equation}
  N_{{\bf k}}=F_{1}(\infty)\Phi_{2}(k), \quad  \chi_{{\bf k}}=
   F_{1}(\infty)\Phi_{1}(k).
       \label{etahi}     \end{equation}
Here, $N_{{\bf k}}$ is the amount of atoms with momentum ${\bf k}$. For $k=0$,
we have $\varphi_{2}(r)\rightarrow 0$ as $r\rightarrow\infty$, and
$\Phi_{2}(0)=N$; therefore, $N_{{\bf k}=0}=NF_{1}(\infty)$, and we obtain the
one-particle condensate $N_{1}\equiv N_{{\bf k}=0}$. In (\ref{w22}) and
(\ref{etahi}), $F_{2}(\infty)=F_{1}^{2}(\infty)$, and $F_{i}(\infty)$
correspond to the 1S-approximation: $\frac{N_{1}}{N} \equiv
F_{1}(\infty)=e^{I_{1A}}$, see (\ref{N0}), (\ref{I1A}).

 As it can be seen from (\ref{wris}), $W_{{\bf k},-{\bf k}}$ consists
 of two terms: the first one, $\frac{N_{{\bf k}}^2}{N^2}$,
 is simply the product of relative numbers of atoms with momenta
 ${\bf k}$ and $-{\bf k}$, this term does not describe
 correlations. The second term,   $\frac{|\chi_{{\bf k}}|^2}{N^2}$,
 describes correlations in ${\bf k}$-space in the pairs $({\bf k},-{\bf k})$.
  It is naturally to relate the two-particle
  condensate precisely with the correlation term $\frac{|\chi_{{\bf k}}|^2}{N^2}$.

   The 2PC was already calculated by Ristig \cite{ris} using a different method.
   In Ristig's works \cite{ris}, the 2PC was determined by the quantity
 \begin{equation}
 P_{2} = \frac{\sum\limits_{{\bf k}\neq 0}\chi_{{\bf k}}^2}
 {\sum\limits_{{\bf k}\neq 0}\left (\chi_{{\bf k}}^2+N_{{\bf k}}^2
 \right )}.
        \label{P2}     \end{equation}
              \begin{figure}
 \includegraphics[width=0.47\textwidth]{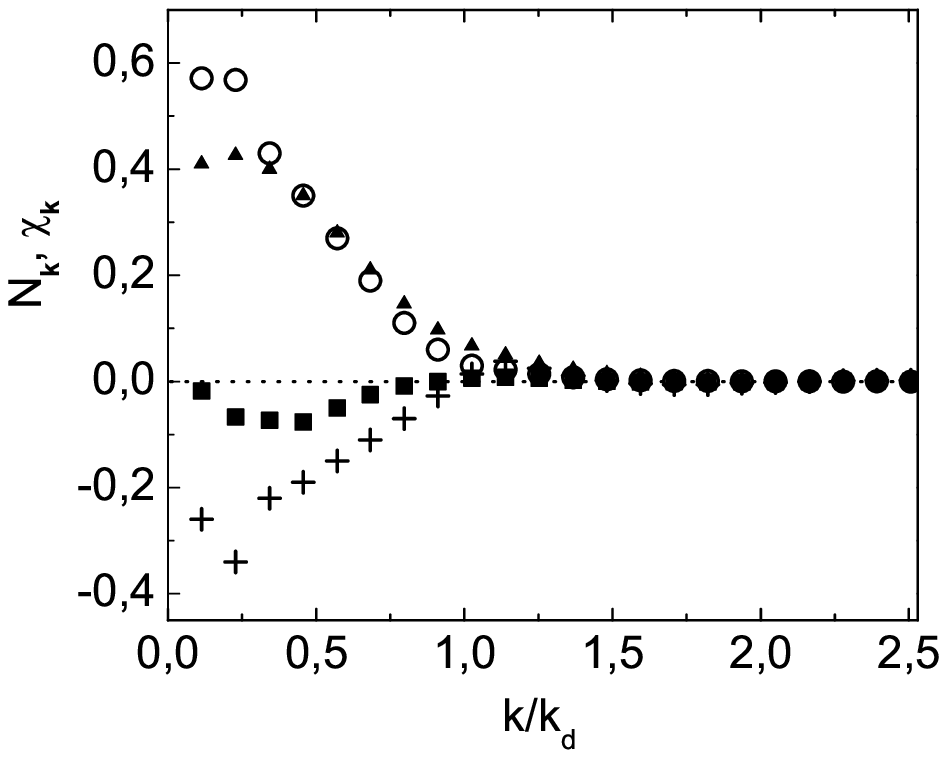}
\hfill
 \includegraphics[width=0.47\textwidth]{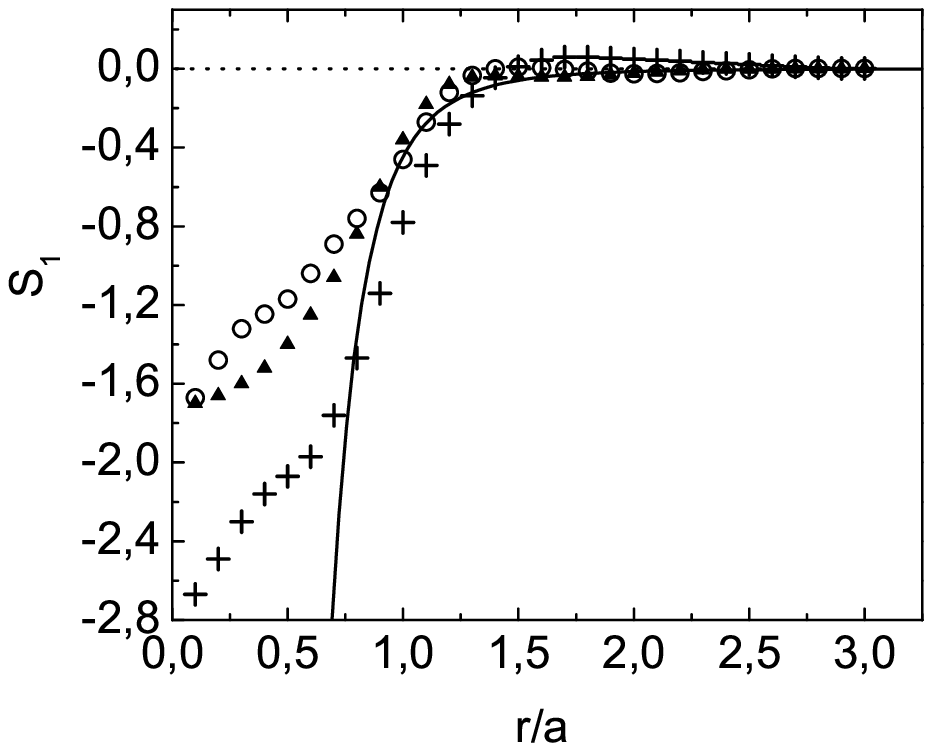}
 \\
 \parbox[t]{0.47\textwidth}{
 \caption{The functions $N_{{\bf k}}$  and $\chi_{{\bf k}}$, where
 $k$ is in the units of $k_{d}=2\pi/d=1.756\AA^{-1}$. Open circles
 mark $N_{{\bf k}}$ in the 1S-approximation; triangles
 mark $N_{{\bf k}}$ in the 2S-approximation; crosses
 mark $\chi_{{\bf k}}$ in the 1S-approximation; squares
 mark $\chi_{{\bf k}}$ in the 2S-approximation.} } \hfill
 \parbox[t]{0.47\textwidth}{
 \caption{The function $S_{1}(r/a)$ (\ref{S11}), where $a=2.64\AA$
 is the diameter \cite{aziz79} of a He$^4$ atom. Open circles
 correspond to the 1S-approximation for $a_{2}(k)$; crosses
  to the 2S-approximation for $a_{2}(k)$; triangles  to the
  ``three sums'' approximation for $a_{2}(k)$ in the approach of
   \cite{jetp} with a model elliptical potential with $U(r=0)=60K$;
  continuous line shows the McMillan's form  (\ref{S1ris}).} }
 \end{figure}
The expression  $P_{2}$ can be interpreted  as the mean degree of
correlation of the pairs of atoms with momenta ${\bf k}_{1}+{\bf
k}_{2}=0$, ${\bf k}_{i}\neq 0$.   We believe that it is more
reasonable to define the {\it number\/} of atoms in 2PC as
follows:
      \begin{equation}
  N_{2} = \sum\limits_{{\bf k}\neq 0}N_{{\bf k}}c_{{\bf k}},
         \label{N2}     \end{equation}
 where $c_{{\bf k}}$ is the ``correlation factor'',
\begin{equation}
  c_{{\bf k}} =
\left [ \begin{array}{ccl}
   \frac{\vert\chi_{{\bf k}}\vert}{N_{{\bf k}}}    & \mbox{for}
&\vert\chi_{{\bf k}}\vert < N_{{\bf k}}, \\
   1  & \mbox{for}
&\vert\chi_{{\bf k}}\vert \geq N_{{\bf k}}.
   \label{ck} \end{array} \right. \end{equation}
  Using $a_{2}(k)$ in (\ref{a2o}), from  (\ref{fi})--(\ref{etahi}) we obtain
 the functions $\chi_{{\bf k}}$ and $N_{{\bf k}}$ (see Fig.~1); then from
 (\ref{P2})--(\ref{ck}) we find $N_{2}\approx 0.53N$ and $P_{2} \approx 0.31$.
In  \cite{ris}, using the Jastrow approximation (\ref{jast}) for $\Psi_{0}$ [as
in our work (\ref{osnj})], it was found $P_{2} \approx 0.09$, but, for
 $S_{1}(r)$ in  (\ref{jast}), Ristig applied \cite{ris} the popular  McMillan's form
   \begin{equation}
 S_{1}(r)=-\frac{r_0^5}{4r^5}, \quad r_{0}=2.963\,\mbox{\AA}.
  \label{S1ris}  \end{equation}
  \begin{figure}[h]
\centerline{\includegraphics[width=85mm]{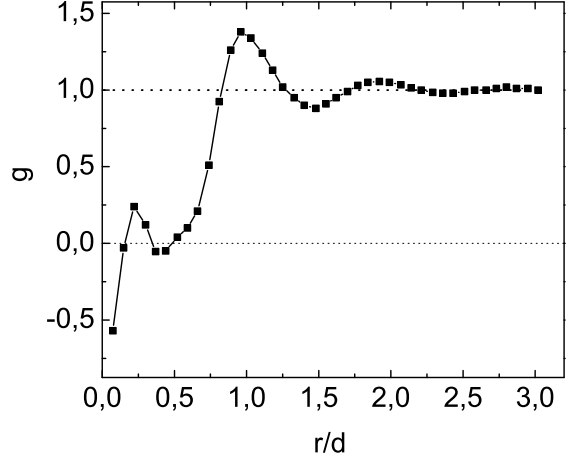}}
\caption{The pair distribution function $g(r)$ of He-II at $T=0$,
from (\ref{g}).}
\end{figure}
    This form is simple and is convenient for calculation but, at the same
 time, it is a very crude variational approximation.
  Using the CV-method, we find $S_{1}(r)$ (see Fig.~2) significantly more accurately
   from (\ref{a2}) and (\ref{S11})
  without any free parameters.  Equation (\ref{a2}) was deduced \cite{vak2,fnt}
  from the equation
 \begin{equation}
S(k)=1+\frac{N}{V}\int d{\bf r}(g(r)-1)e^{-i{\bf k}{\bf r}},
\label{g}  \end{equation}
 connecting $S(k)$ with the pair distribution function $g(r)$.
 Equation (\ref{a2}) was derived in the approximation
neglecting $a_{n\geq 4}$ in  (\ref{osn}) and neglecting multiple
 scattering in  (\ref{g}) \cite{pj}. As shown in Fig.~2,
 our result $S_{1}(r)$ differs greatly from (\ref{S1ris}); therefore,
 our result for $P_{2}$ differs from that obtained in \cite{ris} although the
 formulas for $\chi_{{\bf k}}$ [see (\ref{chiris}) below] and $P_{2}$ are the same
 in our work and in \cite{ris}.
 Note that
 $S_{1}(r\rightarrow\infty)\sim \frac{1}{r^2}$ since
 $a_{2}(k\rightarrow 0)\sim \frac{1}{k}$, see (\ref{a2}).

 It should also be noted that we found the function $S_{1}(r)$
 not quite presicely at $r \lsim d_{c} \approx 2\,\AA$ (at such $r$, two atoms
 penetrate each other). This is evident from the fact that the function
 $g(r)$ recovered from Eq.~(\ref{g})  oscillates irregularly at
 $r \lsim 1.7\,\AA$ (see Fig.~3) and is different from that expected
 from  physical considerations:
 $g(r)\rightarrow 0$ at $r \lsim d_{c}$, and $g(r)>0$. Inaccuracy of the determination
 of $g(r)$ and $S_{1}(r)$ at $r \lsim d_{c}$ is caused by
 large relative error of measurement of the value $S(k)-1$ at $k \gsim 2\pi/d_{c} \approx
  3\,\AA^{-1}$, probably by neglecting multiple
 scattering in  (\ref{g}) \cite{pj},
  and by truncation of expansion (\ref{osn}) [for $S_{1}(r)$].

 Oscillations  of $g(r)$ at small $r$ arouse mainly \cite{ss} from the inaccuracy of measurement
of $S(k)$ at large $k$, $k \gsim 2\pi/d_{c}$. Although the values
of $S(k)$ are very close to unity at such $k$, $S(k)\approx 1$,
the values of $g(r)$ are sensitive to the small oscillations of
$S(k)$ in the neighborhood of the unity. At the same time, such a
small indefiniteness of $S(k)$ at large $k$ influences
insignificantly the amounts of the condensates and quasiparticle
spectrum \cite{ujp}. According to our numerical analysis, the
amount of the condensates are sensitive, first of all, to the
value of $S(k)$ at the small and  middle $k$, $k \lsim 2\pi/d
\approx 1.7\,\AA^{-1}$ (where $d=3.58\AA$
   is the mean distance between helium atoms),
    which corresponds to $r\gsim d$ ($g(r)$
is well defined at such $r$). In fact, the condensates are
``spread out'' in the whole system, so the values of $g(r)$ at
{\em large\/} $r$, $r\gsim d$, are important for the estimates of
the amount of condensates. Moreover, the oscillations of $g(r)$ at
small $r$ may be related to the problem of realistic description
of the structure of helium atoms.

Correlations in the pairs ($0,0$) are absent   \cite{fnt}; therefore, only
atoms with $k>0$ belong to the two-particle condensate $({\bf k},-{\bf k})$.

    In the more exact 2S-approximation, the two-particle condensate is
    calculated similarly (see \cite{fnt} for more details). The function $W_{{\bf k},-{\bf k}}$
 is again defined by (\ref{w22})--(\ref{etahi}) with $\varphi_i$ given by (\ref{phi}),
 where $F_{i}(\infty)$ now correspond to the 2S-approximation [see (\ref{N0})--(\ref{I2C})
 for $\frac{N_{1}}{N}\equiv F_{1}(\infty)$], and we derive more exact
 formulas for $f_1(k)$ and $f_2(k)$:
\begin{equation}
   f_{1}(k)=a_{2}(k)+f_{2}(k)+\delta_{1}(k),
   \label{f1new}     \end{equation}
   \begin{equation}
   \delta_{1}(k)=\frac{1}{N}\sum\limits_{{\bf q}\neq 0}
   \frac{a_{3}({\bf k},{\bf q})}{1-2a_{2}({\bf k}+{\bf q})},
     \label{d1}     \end{equation}
 \begin{equation}
      f_{2}(k)=\frac{a_{2}^{2}(k)}{1-2a_{2}(k)}+\delta_{2}(k)+\delta_{3}(k),
     \label{f2new}     \end{equation}
\begin{equation}
   \delta_{2}(k_{1})=\frac{1}{N}\frac{1}{1-2a_{2}(k_{1})}
   \sum\limits_{{\bf k}_{2}\neq 0}
   \prod\limits_{j=1}^{3} \frac{a_{2}(k_{j})}{1-2a_{2}(k_{j})},
        \label{d2}     \end{equation}
 \begin{eqnarray}
   &&\delta_{3}(k_{1})=\frac{1}{N}\sum\limits_{{\bf k}_{2}\neq 0}
    \left (\prod\limits_{j=1}^{3} \frac{1}{1-2a_{2}(k_{j})}\right )
  a_{3}({\bf k}_{1},{\bf k}_{2}) \\ &&
\times   \left [\frac{a_{2}(k_{1})+a_{3}({\bf k}_{1},{\bf k}_{2})\left (1 -
  a_{2}(k_{1})\right )}{1-2a_{2}(k_{1})} +a_{2}(k_{1})\left (1-2a_{2}(k_{2})
  \right )+a_{2}(k_{2})a_{2}({\bf k}_{1}+{\bf k}_{2}) \right ].\nonumber
        \label{d3}     \end{eqnarray}
   The function $\chi_{{\bf k}}$ in the
   2S-approximation  can be presented in the form
    \begin{equation}
  \chi_{{\bf k}} = \frac{N}{V}\int d{\bf r}
  e^{2S_{1}(r)+S^{*}_{2}(r)}F_{1}(r)e^{i{\bf k}{\bf
  r}}, \quad k\neq 0,
    \label{chiris}     \end{equation}
  where $F_{1}(r)=\frac{N_1}{N}e^{\varphi_{2}({\bf r})}$
  (this is the one-particle density matrix in the 2S-approximation, see Fig.~4),
   $S_{1}(r)$ is defined according to (\ref{S11}), and
      \begin{equation}
 S^{*}_{2}(r) = \frac{1}{N}\sum\limits_{{\bf k}\neq 0}\delta_{1}(k)
 e^{i{\bf k}{\bf r}}.
   \label{S2*}    \end{equation}
            \begin{figure}
 \includegraphics[width=0.47\textwidth]{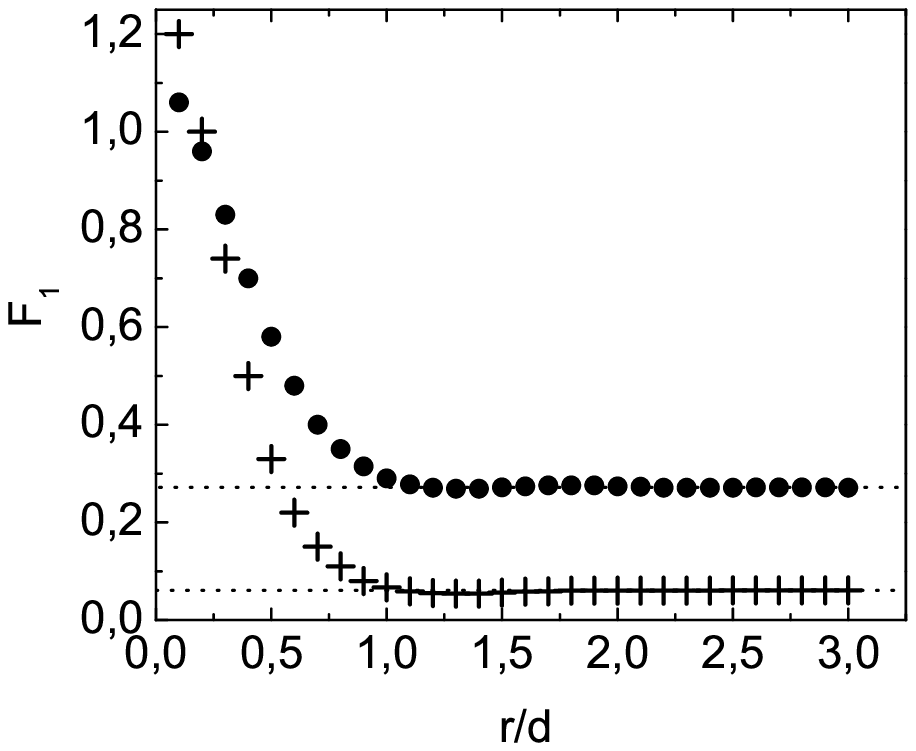}
 \hfill
 \includegraphics[width=0.47\textwidth]{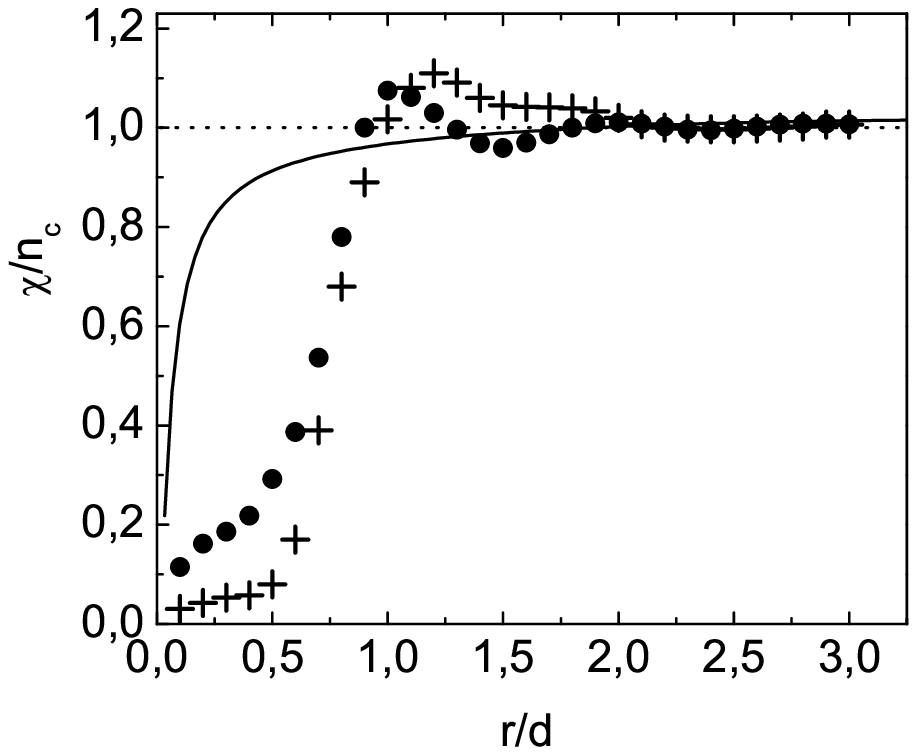}
 \\
 \parbox[t]{0.47\textwidth}{
 \caption{The one-particle density matrix $F_{1}(r)$ for He-II.
 Circles  correspond to 1S-approximation,
   crosses correspond to 2S-approximation.
  The dotted line shows the amount of one-particle
  condensate,  $\frac{N_{1}}{N}=F_{1}(\infty)$, for the
  approximations of 1S and 2S.} } \hfill
 \parbox[t]{0.47\textwidth}{
 \caption{The function $\chi(r)/n_{c}$ (\ref{chir}), where
 $d=3.578\,\AA$
 is the mean distance between He-II atoms, and $n_{c}=N_{1}/V$.
  Circles  and crosses correspond to the same approximations as in
  Fig.~4; the continuous line shows the approximate form
 of the functions $g(r)$ and $\chi(r)/n_{c}$ for a weakly interacting Bose
 gas \cite{fnt}; the dotted line indicates the level  $\chi(r)/n_{c}=1$.} }
 \end{figure}


  For the 1S-approximation, we have $\delta_{1}(k)\equiv 0$, $S^*_2(r)=0$, and
  $F_{1}(r)=F_{1}^{(1)}(r)$, so that (\ref{chiris})  coincides with the
   Ristig's formula for $\chi_{{\bf k}}$ derived earlier \cite{ris}
    (to an accuracy of the factor $N/V$ appearing as a result
  of different normalization of $F_1$; we also choose the opposite sign for
   $\chi_{{\bf k}}$).  Formula   (\ref{chiris}) was not obtained previously.

  Using  $a_{3}({\bf k}_{1},{\bf k}_{2})$ given by (\ref{a3}),
  and   $a_{2}(k)$ as a solution of (\ref{a2}), we find the
  two-particle condensate according to (\ref{etahi}--\ref{N2}),
   (\ref{f1new})--(\ref{d3}): $N_{2}\approx 0.16N$,
  $P_{2} \approx 0.025$.   The obtained functions $\chi_{{\bf k}}$
  and $N_{{\bf k}}$ are shown in Fig.~1.

 Note that, in the 2S-approximation, we have, taking into account (\ref{a20}),
  $N_{{\bf k}\rightarrow 0} =
 \frac{N_{1}}{N}\frac{a_{2}(k)}{2}=\frac{N_{1}}{N}\frac{mc}{2\hbar k}
 \left (1+\Sigma_{2}(0) \right )$,
 $\chi_{{\bf k}\rightarrow 0} =-N_{{\bf k}\rightarrow 0}$.
  Such an asymptotics for $N_{{\bf k}}$ refines the one
   found earlier in \cite{gn,ches}, $N_{{\bf k}\rightarrow 0} =
   \frac{N_{1}}{N}\frac{mc}{2\hbar k}$.

    In Fig.~5, we show the function
  \begin{equation}
  \chi(r)=\frac{1}{(2\pi)^3}
  \int d{\bf k}e^{-i{\bf k}{\bf r}}\chi_{{\bf k}}=
  \frac{N}{V}e^{2S_{1}(r)+S^{*}_{2}(r)}F_{1}(r)=
   \frac{N_1}{V}e^{\varphi_{1}({\bf r})}.
   \label{chir}    \end{equation}
  The function $g(r)$ describes the dependence  of the correlations
  in the pairs of two atoms with arbitrary momenta, (${\bf k}_1, {\bf
  k}_2$), on the size of the pair $r$; and $\chi(r)$ describes the
   same for a narrower class of pairs (${\bf k},-{\bf
 k}$). For infinite separation between the atoms in the pair, correlations
  disappear; therefore, $g(\infty)$ and $\chi(\infty)$ take the
  values of $g$ and $\chi$ for uncorrelated pairs [for a totally
  chaotic distribution of atoms in the ${\bf r}$-space, we would
  have $g(r)\equiv g(\infty)$ for all $r$, and similarly for $\chi(r)$].

  Comparing the functions $g(r)$ (Fig.~3) and  $\chi(r)$ (Fig.~5),
  we can see that these functions have similar form, as it could be
   expected.
  Atoms interact strongly at a small distance, so the correlations
  in the pares are significant at $r\lsim 2d$, where the values of
  $g(r)$ and  $\chi(r)$ are  substantially different from those at
  infinity.

   The authors of \cite{ris+,pif} believe to be warrant the investigations
   of the possible pairing of He-II atoms in the ${\bf r}$-space. However,
   in our opinion, the properties of  $\chi(r)$ and the results of \cite{ris+}
   do not give grounds to talk about the bound states in the ${\bf
   r}$-space; these properties are indicative only of
correlations in pairs.
 The pairing energy  $E_{p} \approx -0.05$~K was found in \cite{ris+},
 but this energy is too small for the appearance of real bound pairs
 because $E_{p}$ is two orders of magnitude smaller than the ground-state energy $E_{0}$.

  At present, we are not aware of any reliable evidence of the existence
  of bound pairs of atoms in  He-II. An interesting arguments in favour of the
  possibility
  of pairing were presented in \cite{pif}.  Our analysis cannot exclude the possibility
  of existence of {\it narrow\/} bound pairs, of atomic size.
  Electrons should be collectivized in narrow pairs; therefore, such
  pairs have to be considered as a new kind of ``atoms''; in this
  case, we need to know
  $\Psi_{0}$ for a mixture of two Bose liquids. However, the closeness
   of the experimental value of circulation \cite{cir} to the value of one
  quantum, $\kappa=\hbar/m$, indicates that the number of bound pairs in
   He-II is very small or equal to zero.

  \section{The higher condensates}
  We also investigated correlations (in the ${\bf k}$-space)
  in groups of  $s$ atoms  ($s\geq 3$) under the condition
   ${\bf k}_{1}+\ldots +{\bf k}_{s}=0$ for momenta  ${\bf k}_{i}$ of atoms.
   We were interested in the ``nontrivial'' part of the $s$-particle condensate,
  which cannot be reduced to the ``lower''  condensates; so we put
   ${\bf k}_{1}+\ldots +{\bf k}_{l}\neq 0$ for all $l=1,\ldots,s-1$.
   There is a wide-spread, but not proved, opinion according to which the higher
   condensates {\it should be present\/} in He-II. We considered the
   problem in the 2S-approximation, for which the three-particle term
   with $a_3$ is taken into account in  $\ln{\Psi_{0}}$ (\ref{osn}),
   and we expected that at least three-particle condensate would turn out to be nonzero.
   However, we found \cite{fnt} that all  higher s-particle condensates ($s\geq 3$)
   {\it are absent\/} from He-II at $T=0$.
   An analysis is cumbersome and similar to that for 2PC, see \cite{fnt}
   for more detail.

      Our results, which consist in the absence of all  higher s-particle condensates ($s\geq 3$)
      and in the small value of two-particle condensate,  should be important for the
   field-theoretic model of He-II: they can significantly simplify such
   model and, perhaps, open a possibility of constructing an ``ideal''
    micro-model of He-II.

   \section{Conclusions}

The main results of our paper were already outlined in the
introduction. Summarizing, we notice that the CV-method allows one
to calculate the quasiparticle spectrum of   He-II
\cite{yuvdan,ujp,jetp,megot} and also, with lower accuracy,
condensates (this work), without any fitting parameters or
functions. The equations of the model are deduced from the exact
microscopic equations, and the results approximately agree with
the experiment. Thus, the model under consideration, undoubtedly,
gives an approximate description of the microstructure of He-II.

   However, the calculated amount of one- and two-particle condensates
   appreciably depends on the number of the corrections $a_n$ taken into
   account in $\ln{\Psi_{0}}$. The reason is that the quantity $\ln{\Psi_{0}}$ and
  the condensates $N_i$ are expanded in series in a parameter which
  is not very small (this is the function  $q(k)=a_{2}(k)k/k_{d}$, $k_d=2\pi/d$,
  where $d$
   is the mean distance between atoms; the mean value of $q(k)$
  at $k<k_{d}$ is not very small, being around $-1/2$, see Fig.~1 in
  \cite{fnt}) and, for condensates, the series stand in the exponent. These
  difficulties  arise in all approaches where condensates are
  calculated from $\Psi_{0}$, unless the free parameters are chosen so that
  the corrections ``turn out'' to be small.

If, in some  quantum mechanical approach, the amount of the
condensate changes insignificantly by taking into account the next
corrections to $\ln{\Psi_{0}}$ and the  condensates, this is,
probably, mainly a result of a good choose of fitting parameters.
Really, in our model we find a solution from the exact (but
truncated) microscopic equations, without free parameters
whatsoever, and we clearly see that the corrections to
$\ln{\Psi_{0}}$ and especially to the values of the condensates
are not small enough, unfortunately.
 Nevertheless, our results give the approximate
estimates of the  fraction of the condensates (the amount of 1PC
agrees with the experiment) and, at the same time, the significant
value of the corrections  show that quantum-mechanical approaches
(except, perhaps, the MC methods) do not offer methods for
sufficiently precise calculation of the amount of the condensates.

   The Monte Carlo method \cite{qdmc,mor} is one of the  perspective approaches  for
 the  calculation of the 1PC and 2PC but, unfortunately, we do not know
   the  error of the numerical definition of 1PC and 2PC for the
   MC simulations.

The absence of small parameter is a general problem of virtually
all (except the MC simulations) known approaches  to the
description of the microstructure of He-II\@. Perhaps, in some
approaches, in the future, one will succeed in calculating the
1PC, 2PC and higher condensates more exactly, using expansions in
small parameters only and without fitting parameters and
unjustified postulates. Unfortunately, by now, such an ``ideal''
micromodel of He-II is not constructed.  And it is not clear even,
whether it is possible.

   The author is very
grateful to Yurii V. Shtanov for valuable discussion.

\renewcommand\refname{REFERENCES}

        \end{document}